\documentclass[twocolumn,pre,aps,showpacs,amsmath,amssymb,superscriptaddress]{revtex4}

\usepackage{hyperref}
\usepackage{amsmath,amssymb}
\usepackage{amsfonts,amsthm}
\usepackage{graphics}
\usepackage{graphicx}
\usepackage{dcolumn}
\usepackage{color}
\usepackage{bm}
\usepackage[dvipsnames]{xcolor}
\usepackage[normalem]{ulem}
\usepackage[bf]{subfigure}

\begin{document}

\title{
Statistical complexity is maximized close to criticality in cortical dynamics\\
}

\author{Nastaran Lotfi}
\affiliation{Departamento de F\'{\i}sica, Universidade Federal de Pernambuco, Recife PE 50670-901, Brazil.}

\author{Tha\'{\i}s Feliciano}
\affiliation{Departamento de F\'{\i}sica, Universidade Federal de Pernambuco, Recife PE 50670-901, Brazil.}

\author{Leandro A. A. Aguiar}
\affiliation{Departamento de Ci\^encias Fundamentais e Sociais, Universidade Federal da Para\'iba, Areia PB 58397-000 Brazil.}

\author{Thais Priscila Lima Silva}
\affiliation{Departamento de F\'{\i}sica, Universidade Federal de Pernambuco, Recife PE 50670-901, Brazil.}

\author{Tawan T. A. Carvalho}
\affiliation{Departamento de F\'{\i}sica, Universidade Federal de Pernambuco, Recife PE 50670-901, Brazil.}


\author{Osvaldo A. Rosso}
\affiliation{Instituto de F\'{\i}sica, Universidade Federal de Alagoas, Macei\'{o}, Alagoas 57072-970 Brazil.}

\author{Mauro Copelli}
\affiliation{Departamento de F\'{\i}sica, Universidade Federal de Pernambuco, Recife PE 50670-901, Brazil.}

\author{Fernanda S. Matias}
\thanks{fernanda@fis.ufal.br}
\affiliation{Instituto de F\'{\i}sica, Universidade Federal de Alagoas, Macei\'{o}, Alagoas 57072-970 Brazil.}

\author{Pedro V. Carelli}
\thanks{pedro.carelli@ufpe.br}
\affiliation{Departamento de F\'{\i}sica, Universidade Federal de Pernambuco, Recife PE 50670-901, Brazil.}

\begin{abstract}
Complex systems are typically characterized as an intermediate situation between a complete regular structure and a random system. Brain signals can be studied as a striking example of such systems: cortical states can range from highly synchronous and ordered neuronal activity (with higher spiking variability) to desynchronized and disordered regimes (with lower spiking variability). It has been recently shown, by testing independent signatures of criticality, that a phase transition occurs in a cortical state of intermediate spiking variability. Here, we use a symbolic information approach to show that, despite the monotonical increase of the Shannon entropy between ordered and disordered regimes, we can determine an intermediate state of maximum complexity based on the Jensen disequilibrium measure. More specifically, we show that statistical complexity is maximized close to criticality for cortical spiking data of urethane-anesthetized rats, as well as for a network model of excitable elements that presents a critical point of a non-equilibrium phase transition.

\end{abstract}
\maketitle

%
%

\section{Introduction}
Complexity is a ubiquitous concept in modern science and life. 
A complex dynamical system is typically associated with a mixture of order and disorder as well as to emergent phenomena, often across multiple temporal and spatial scales. 
Although a universal and precise definition of complexity is still lacking, different measures for complexity have been proposed in the literature: Kolmogorov's complexity, 
an algorithmic information content based on the size of the smallest computer program that can produce an observed pattern~\cite{Kolmogorov65,Chaitin77}; Crutchfield and Young’s complexity~\cite{Crutchfield89}, which measures the amount of
information about the past required to predict the future; and
a measure of the self-organization capacity of
a system~\cite{Sprott03}.

The Martín-Platino-Rosso (MPR) statistical complexity~\cite{Martin06} employed here is evaluated using the Bandt-Pompe~\cite{Bandt02} recipe to assign a probability distribution function to the time series generated by the system of interest. It is based on the Jensen disequilibrium measure and tends to zero for both perfectly regular and random signals. Since the normalized Shannon entropy goes from zero to one between those extremes, the multi-scale entropy–complexity causality plane is a useful tool to characterize complex systems~\cite{Martin06}. Moreover, the method allows us to evaluate the complexity at different time scales by using a symbolic information approach for different down-samplings.

The multi-scale complexity-entropy causality plane has been also used to identify the range of scales at which nonlinear deterministic or stochastic behaviors dominate the system’s dynamics~\cite{Rosso07,Zunino12}.
In neuroscience, it has been employed to estimate the time delay between synchronized cortical areas of a non-human primate during a cognitive task~\cite{Montani15}, in neuronal network descriptions~\cite{Montani2014,Montani2015neuronas} as well as in EEG signals during epileptic seizures~\cite{Rosso06}.

Recently, consistent markers of a phase transition in brain signals have been reported at an intermediate level of neuronal spiking variability between a synchronized (ordered) state and a desynchronized (disordered) state~\cite{Fontenele19,lotfi2020signatures}.
It has been proposed that, since the cortex operates in both extreme modes during different cognitive functions~\cite{hasselmo1995neuromodulation}, it could be advantageous to self-organize close to the critical point between them.
A link between criticality and complexity was proposed by Timme et al.~\cite{Timme16}, whose analysis of neuronal avalanches have revealed that cortical branching models exhibit a local peak in complexity close to the critical point.
Moreover, they have shown that complexity in culture data is larger than for randomized neuron identities data, which supports the hypothesis that  complexity is maximized near the critical point.

Here we apply a symbolic information-theoretical approach to neuronal firing rate time series to quantify the permutation entropy and MPR complexity across the full range of recorded cortical states. 
In Sec.~\ref{methods}, we describe the symbolic information approach and the information theory metrics employed in our data analysis. 
In Sec.~\ref{results}, we report our results, showing that the statistical complexity is maximized close to the critical point between synchronized and desynchronized cortical states for both urethane-anesthetized rats and a network model presenting a non-equilibrium phase transition.
Concluding remarks and a brief discussion of the significance of our findings for neuroscience are presented in Sec.~\ref{conclusions}.


\section{\label{methods} Methods}

\subsection{Information-theoretical quantifiers}

An information measure can be viewed as a quantity that characterizes some property of a given probability distribution function (PDF). 
To calculate any information-theory quantifier, one should obtain a PDF from a time series $X(t)$ representing the evolution dynamics of the system under study. 
Let $X(t)\equiv \{x_t;t=1,2, \dots ,M\}$, be the time series representing a set of $M$ measures of the observable $X$. 
It is possible to associate to $X(t)$, by a symbolic information approach described below, a probability distribution function given by $P\equiv \{p_j;j=1,2, \dots ,N\} $ with $\sum_{j=1}^{N} p_j=1$ where $N$ is the number of possible states of the system. Therefore, Shannon’s logarithmic information measure is defined by~\cite{Shannon49}:
\begin{equation}
\label{eq:S}
S[P] = -\sum_{j=1}^{N} p_j \ln(p_j).
\end{equation}

This function is equal to zero when we can correctly predict the outcome every time. 
By contrast, the entropy is maximized for the uniform distribution $P_e = \{p_j=1/N, \forall j=1,2, \dots ,N\}$.
Then, the normalized Shannon entropy is defined by 
$H[P] = S[P] / S[P_e]$ ($0 \leq H \leq 1$).


A complex system cannot be fully characterized only by a randomness measure. The opposite extremes of perfect order and maximal randomness are too simple to describe as they do not have any structure, and the complexity should be zero in both cases. Thus, measures of statistical complexity are needed to gain a better understanding of time series. Here, we consider the MPR statistical complexity~\cite{Lamberti04} as it can quantify critical details of dynamical processes underlying the data-set.
Based on the seminal notion of a statistical complexity based on a disequilibrium advanced by López-Ruiz et al.~\cite{lopez1995statistical}, the MPR statistical complexity measure is defined through the product:
\begin{equation}
\label{eq:C}
C[P] = Q_J[P,P_e] \cdot H[P].
\end{equation}
The disequilibrium $Q_J[P,P_e]$ is defined in terms of the Jensen–Shannon divergence as:
\begin{equation}
\label{eq:Q}
Q_J[P,P_e] = Q_0 J[P,P_e],
\end{equation}
where
\begin{equation}
\label{eq:J}
J[P,P_e] = S\left[\frac{(P+P_e)}{2}\right] - \frac{S[P]}{2} - \frac{S[P_e]}{2},
\end{equation}
and $Q_0$ is a normalization constant ($0 \leq Q_J \leq 1$), equal to the inverse of the maximum possible value of $J[P,P_e]$.
This maximum value is obtained when one of the components of $P$, say $p_m$, is equal to 1, and the remaining $p_j$ are equal to zero.

The Jensen–Shannon divergence $J[P,P_e]$ is a metric to quantify the difference between two probability distributions: $P$ and $P_e$, respectively, the one associated with the system of interest and the uniform distribution. It is especially advantageous to compare the symbolic composition between different sequences~\cite{Grosse02}. 
It has been shown that, for a given value of normalized entropy $H$, the complexity $C$ can vary between a well-defined minimum $C_{-}$ and a maximum $C_{+}$ value, which restricts the possible occupied region in the complexity-entropy plane~\cite{Martin06}. 


\begin{figure}
 \begin{minipage}{8cm}
  \begin{flushleft}(a)%
\end{flushleft}%
\includegraphics[width=0.86\columnwidth,clip]{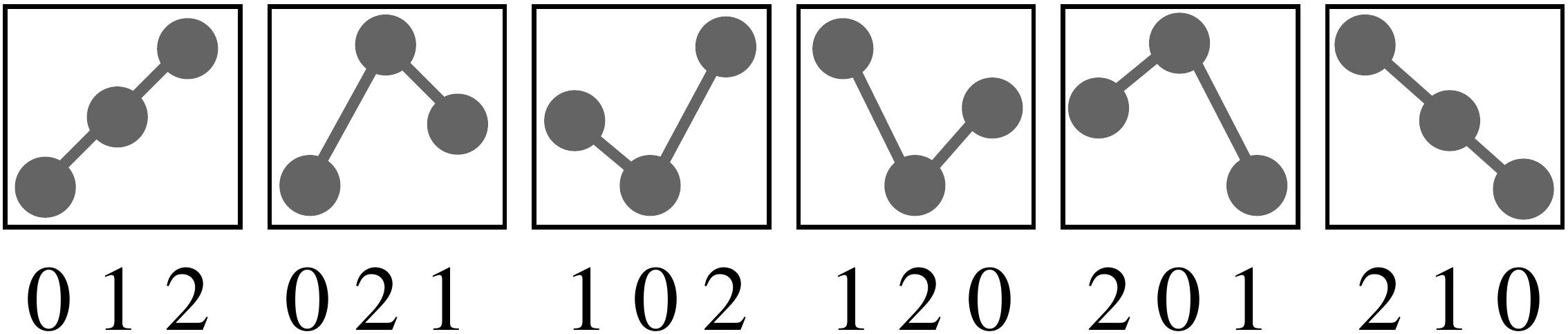}
\end{minipage}
\begin{minipage}{8cm}
\begin{flushleft}%
\end{flushleft}%
\includegraphics[width=0.98\columnwidth,clip]{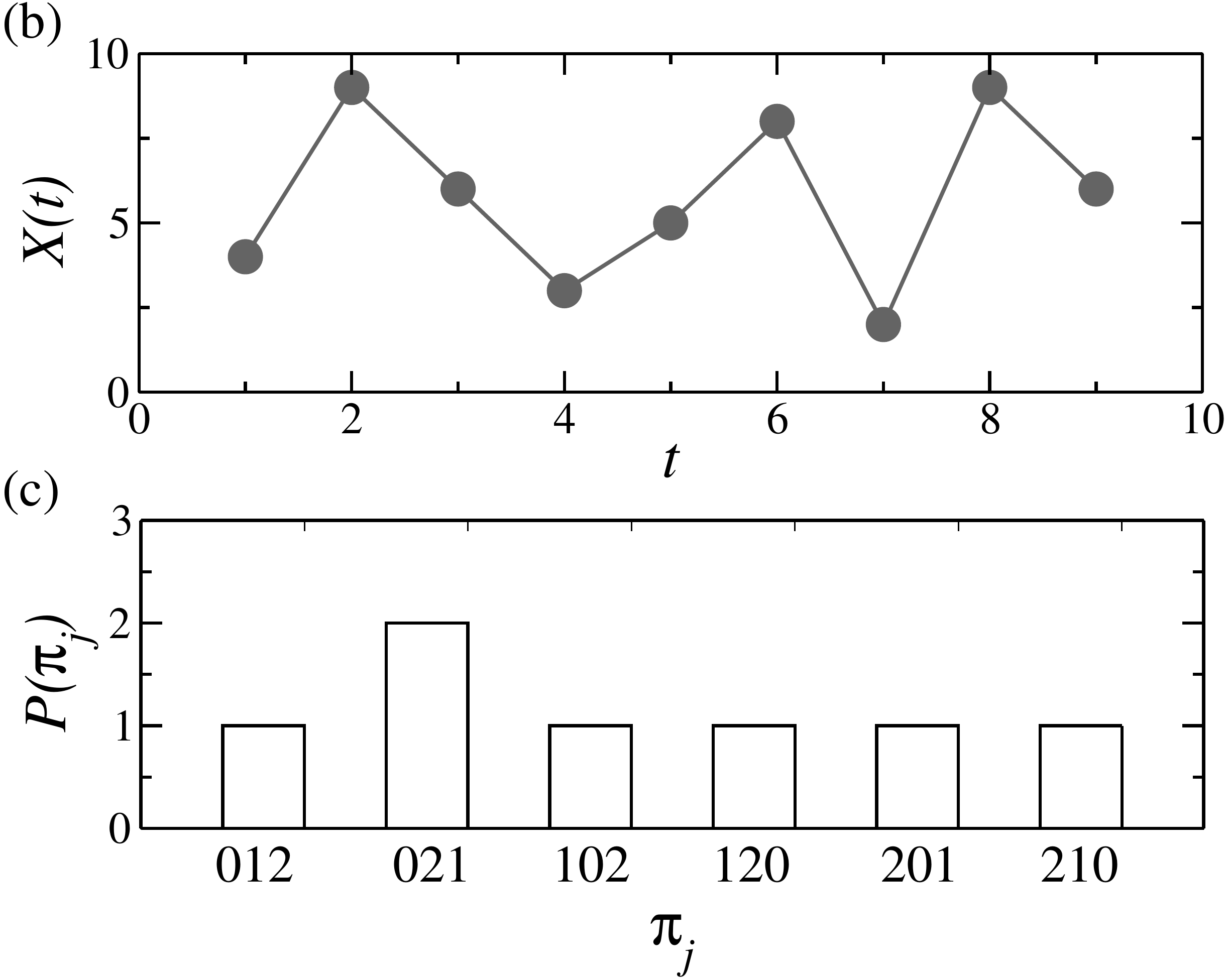}
\end{minipage}
\caption{\label{fig:DSAS} 
Characterizing the symbolic representation of time series.
(a) The six possible symbols associated with permutations $\pi_j$ for ordinal patterns of length $D=3$.
(b) Example of a very simple time series $X(t)$ and (c) its own non-normalized probability density function (PDF).
}
\label{fig:symbol}
\end{figure}

\subsection{Symbolic representation of a time series}

To calculate the two information-theoretical quantifiers mentioned previously, a probability distribution $P$ should be estimated from the time series $X(t)$ of the system. Here, we use the symbolization technique introduced by Bandt \& Pompe (BP)~\cite{Bandt02} for evaluating the PDF, associated with a specific time series. 
We are interested in extracting the ordinal patterns of
length $D$, associated to each time $t$ of our time series, generated by $\textbf{s}(t)=(x_{t-(D-1)},x_{t-(D-2)}, \cdots ,x_{t-1},x_{t})$. This corresponds to indexing each $t$ to the $D$-dimensional vector $\textbf{s}(t)$.
The greater the value of $D$, the more information about the past is incorporated into the vectors. 

We should identify and count the number of occurrences of
all $D!$ permutations $\pi_j$ of length $D$ (with $j=1,2,...,D!)$.
The specific $j-th$ ordinal pattern associated to $\textbf{s}(t)$ is the permutation $\pi_j=(r_{0},r_{1},...,r_{D-1})_j$ of $(0,1,...,D-1)$ which guarantees that $x_{t-r_{(D-1)}} \leqslant x_{t-r_{(D-2)}}  \leqslant  \cdots  \leqslant x_{t-r_{1}}  \leqslant  x_{t-r_{0}}$. 
In order to get a unique result, we set $r_i < r_{i-1}$ if $x_{t-r_i} = x_{t-r_{i-1}}$.
In other words, each permutation $\pi_j$ is one of our possible symbols and we have $D$! different symbols.
Therefore, the pertinent symbolic data is created by the following rules: (i) grouping the $D$ consecutive values of the time series points in the vector $\textbf{s}(t)$, (ii) indexing a symbol $\pi_j$ to the vector $\textbf{s}(t)$ by reordering the embedded data in ascending order using the permutation $\pi_j$. Therefore, for each $x_t$ (with $t=1,2, \dots ,M-(D-1)$), we can associate a symbol $\pi_j$. 

Afterward, it is possible to quantify the diversity of the ordering symbols (patterns) derived from a scalar time series by counting how many times each one of the $D$! different permutations $\pi_j$ have been found in the data-set. Then, to calculate the PDF (for a specific $D$), we find
$P\equiv \{p_j;j=1,2,...,D! \}$, where $p_j$ is the probability to find the $j$-th symbol  $\pi_j$ in our time series. This procedure is essential to a phase-space reconstruction with embedding dimension (pattern length) $D$. For practical purposes, BP suggested to use $ 3\leqslant D \leqslant 7 $.

To have an example, choosing $D=3$, all the 6 possible symbols associated with the permutations $\pi_j$ are presented in Fig.~\ref{fig:symbol}(a). Considering the time series $X(t)=\{4,9,6,3,5,8,2,9,6\}$ as an example (see Fig.~\ref{fig:symbol}(b)), the first vector is $\textbf{s}(t=1)=(4,9,6)$, corresponding to the permutation $\pi_2=(0,2,1)$; the second vector is $\textbf{s}(t=2)=(9,6,3)$, corresponding to to the permutation $\pi_6=(2,1,0)$. Similarly, one can find the other 5 vectors $\textbf{s}(t)$ and its respective $\pi_j$. The correspondent non-normalized PDF is shown in Fig.~\ref{fig:symbol}(c).

Note that the symbol sequences naturally arise from the time series and do not require model-based assumptions. Despite losing some details of the original series’ amplitude, this technique takes into account the temporal structure of the time series and yields information about the temporal correlation of the system [31,32]. Finally, the  BP methodology only requires a very weak stationarity assumption: for $k \leq D$, the probability for $x_t \leq x_t+k$ should not depend on $t$.

To investigate the significance of our results, we compare them to analyses performed on the surrogate data. 
The surrogate data is obtained by randomly shuffling  the interspike intervals of each neuron separately, so that correlations in the original time series are destroyed.

\subsection{Data acquisition}

The experimental firing rates used in our data analysis are taken from two experimental setups as described below: Seven Long-Evans rats, male, 250-360~g, 3-4 months old were used in the recordings. The rats were anesthetized with 1.58 g/kg of fresh urethane, diluted at 20$\%$ in saline, in 3 injections (i.p.), 15 min apart.
Five (two) of the datasets were acquired using 64-(32-)channel silicon probes (BuzsakiA64/BuzsakiA32sp, Neuronexus). These silicon probes are composed of 6 (4) shanks with 10 (8) sites/shank with the impedance of 1-3~MOhm at 1~kHz, in the primary visual cortex of the rats (V1, Bregma: AP $= -7.2$, ML $= 3.5$). Each shank is located with 200~$\mu \text{m}$ distance and each site has an area of 160~$\mu\text{m}^2$, disposed of the tip in a staggered configuration, 20~$\mu \text{m}$ apart. 

Three hours of recorded data were sampled at 30~(24)~kHz, amplified, and digitized in a single head-stage Intan RHD2164 (amplified and digitized in a PZ2 TDT, which transmits to a RZ2 TDT base station). In the last step, using the Klusta-Team software ~\cite{kadir2014high,rossant2016spike} on raw electrophysiological data, we performed spike sorting. From each one of the seven rats we have recorded $N_i$ neurons: 295, 222, 168, 330, 274 for the 64-channel dataset and 130, 146 for the 32-channel dataset. All the experimental procedures were approved by the Federal University of Pernambuco (UFPE) Committee for Ethics in Animal Experimentation (23076.030111/2013-95, 12/2015, and 20/2020).

\begin{figure}[!ht]%
 \begin{minipage}{7cm}
  \begin{flushleft}%
\end{flushleft}%
\includegraphics[width=1\columnwidth,clip]{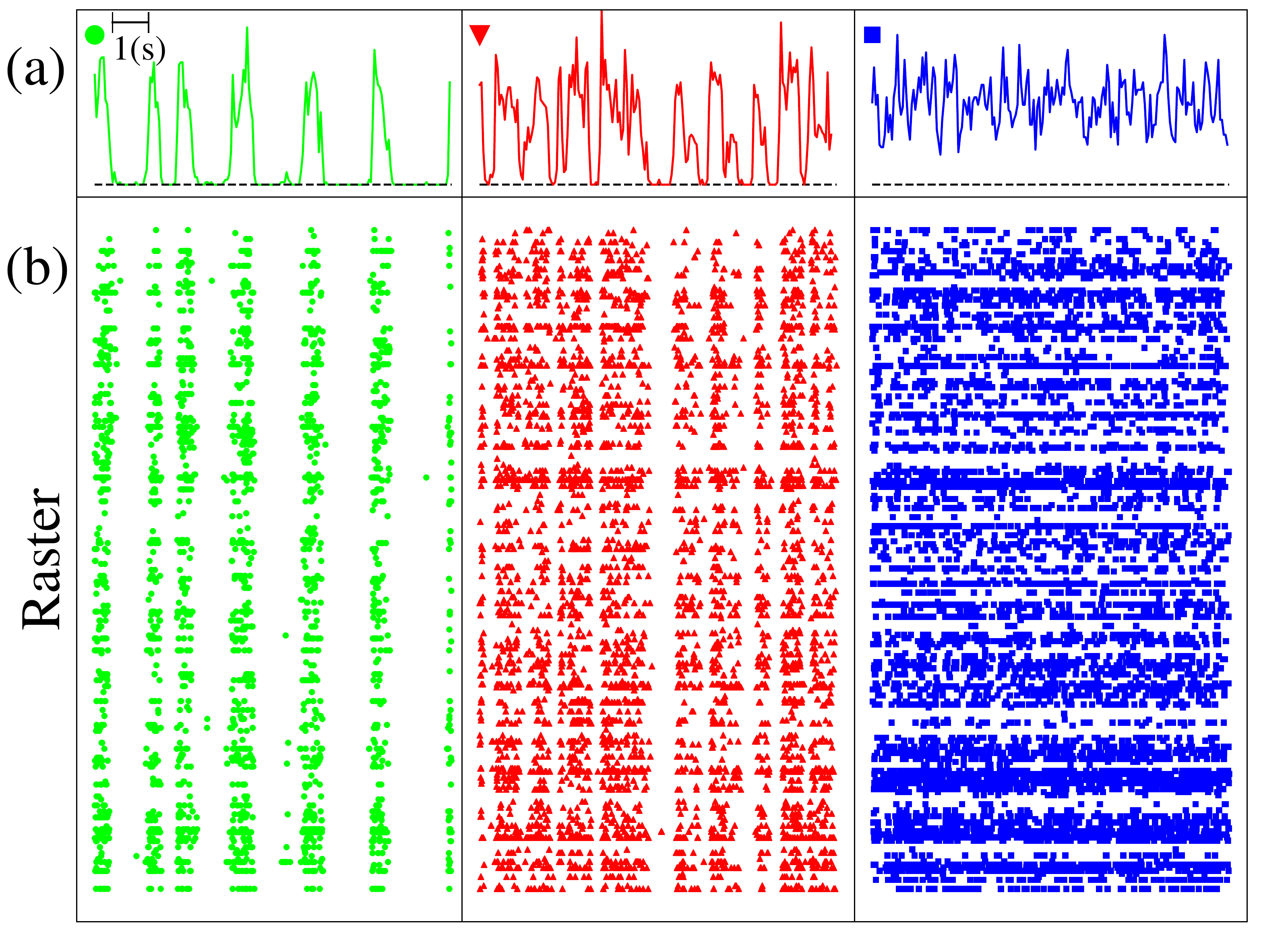}
\end{minipage}
\begin{minipage}{7cm}
\begin{flushleft}%
\end{flushleft}%
\includegraphics[width=1\columnwidth,clip]{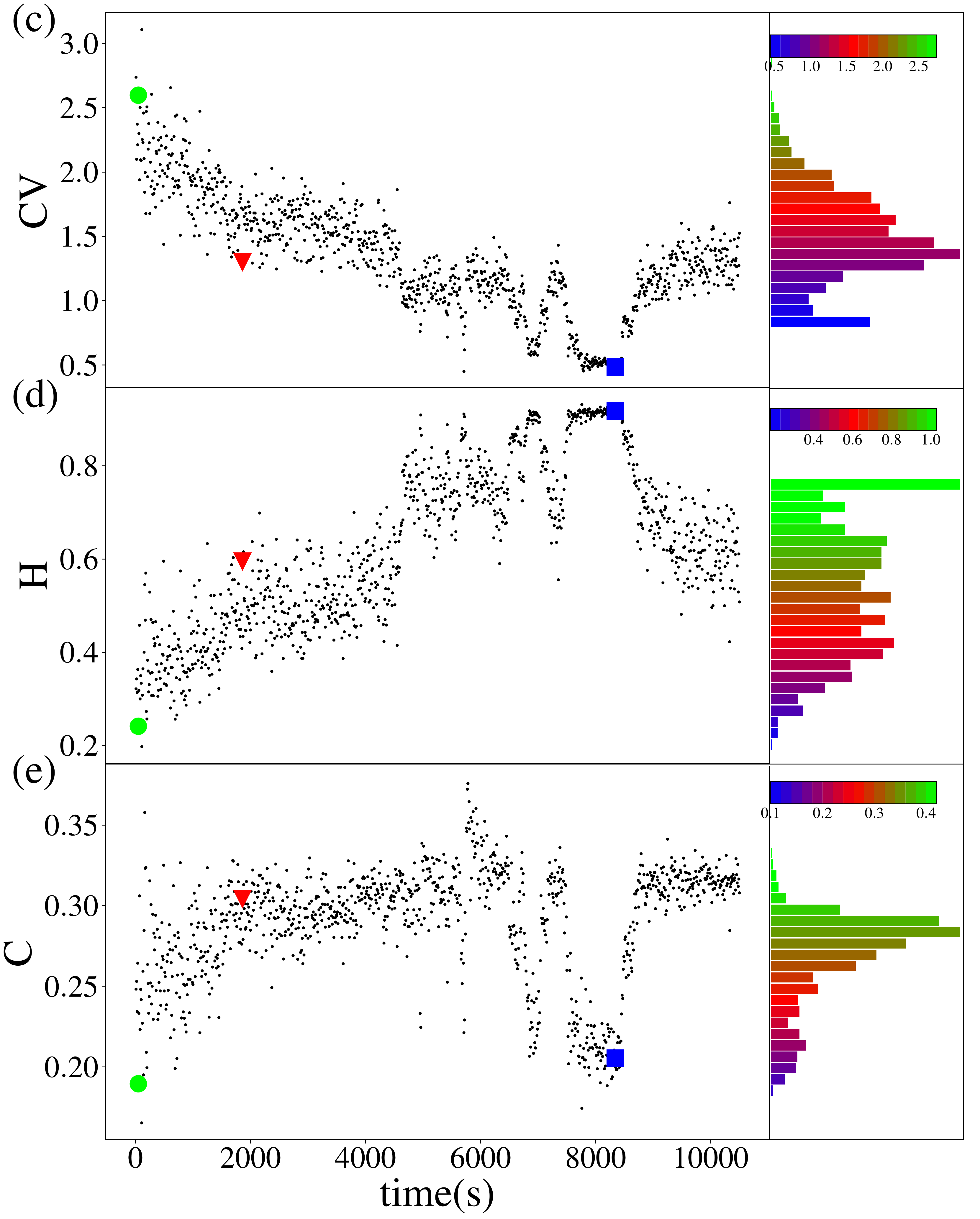}
\end{minipage}
\caption{\label{fig:raster} 
Characterizing cortical states and neuronal variability.
(a) Firing rates $F$ calculated using $10$~ms interval. 
(b) Raster plots for all 274 recorded neurons. (c) Coefficient of variation (CV) of the
spiking activity calculated for $10$-s-long non-overlapping windows of a single animal and the CV histogram. Symbols (circle, triangle, and square) respectively indicate the level of spiking variability of three representative examples: high ($CV=2.60$, highly synchronous and ordered neuronal activity) intermediate ($CV=1.30$), and low ($CV=0.484$, desynchronized and disordered regime.(d) The entropy and (e) the complexity of each window and their histograms. The three illustrative examples exhibit: $H=0.241$ and $C=0.190$ (green circle), $H=0.596$ and $C=0.304$ (red triangle), $H=0.917$ and $C=0.206$ (blue square). 
}
\end{figure}

\section{Results}
\label{results}

\subsection{\label{sec:cv} Characterizing cortical states in anesthetized rats}

For each recorded animal, the data is segmented in windows of duration $W=10$~s, and the firing rate $F$ is computed as the sum of all spikes in time bins of $\Delta t=10$~ms (unless otherwise stated). 
Illustrative examples of the firing rates during three different time windows of a few seconds are shown in Fig.~\ref{fig:raster}(a) with their respective raster plots shown in Fig.~\ref{fig:raster}(b). 
In the left panels, neurons silence together, signalizing a very ordered (synchronized) firing pattern. 
 In the right panels, the silent periods are not present at all, indicating a disordered (desynchronized) regime. 
The middle panels illustrate an intermediate situation between synchronized and desynchronized states.

To better characterize cortical dynamics along the different levels of spiking variability, for each $i$-th window  of duration $W$, its coefficient of variation ($CV_i$) of the population firing rate is calculated as:
\begin{equation}
\label{eq:CV}
CV_i = \frac{\sigma_i}{\mu_i},
\end{equation}
where $\mu_i$ is the mean and $\sigma_i$ is the standard error of $F$ within window $i$. 
We have tested that our results remain qualitatively the same for $W \in [xx~\mbox{s},yy~\mbox{s}]$ and $\Delta t \in [xx~\mbox{ms},yy~\mbox{ms}]$.
In Fig.~\ref{fig:raster}(c), we show how the $CV$ changes during the experiment and the $CV$ distribution for a single rat.

The higher the $CV$, the more synchronized is the cortical state.
In Fig.~\ref{fig:raster}b, left panels show a highly synchronous and ordered neuronal activity with a high spiking variability and long silent periods ($CV=2.60$). Middle panels exhibit an intermediate spiking variability ($CV=1.30$). Right panels illustrate an example of a desynchronized regime with low spiking variability ($CV=0.484$, disordered regime). 
Another way to  frame this connection is by noticing that $CV$ correlates very strongly with the average pairwise correlation of neuronal firing rates~\cite{lotfi2020signatures}.

Previous studies have shown that the cortical dynamics of urethane anesthetized animals hovers around a critical point~\cite{Fontenele19,lotfi2020signatures}. 
By analyzing the distribution of avalanche sizes and lifetimes and a scaling relation connecting the critical exponents, Fontenele et al.~\cite{Fontenele19} have shown that a critical point can be associated to a critical value of $CV=1.4 \pm 0.2$. 
By employing a different analysis based on the Maximum Entropy approach~\cite{Jaynes93}, Lotfi et al have calculated a critical spiking variability value of $CV=1.28 \pm 0.08$~\cite{lotfi2020signatures} for  a similar dataset. 

Here, we propose to characterize the cortical states by means of the symbolic Shannon entropy $H$ described in Sec.~\ref{methods}. 
Starting from the firing rate time series illustrated in Fig.~\ref{fig:raster}(a), we calculate the entropy of each window of width $W$. 
For the three samples highlighted in Fig.~\ref{fig:raster} we obtain: $H=0.241$ (green circle), $H=0.596$ (red triangle), and $H=0.917$ (blue square). 
As expected, the entropy increases from ordered (synchronized) to disordered (desynchronized) states. In  Fig.~\ref{fig:raster}(d), we show how $H$ varies along the experiment as well as its distribution. 
A comparison with the $CV$ time series in Fig.~\ref{fig:raster}b suggests an anti-correlation
between $CV$ and $H$. 
It is worth emphasizing that the calculation of $H$ is independent of the calculation of $CV$. 
In Fig.~\ref{fig:HCCV}(a) we plot all the windows in the ($CV,H$) plane for a single animal, using time bins of $\Delta t=20$~ms. The inverse relation between $H$ and $CV$ shows that the entropy is lower for higher spiking variability.

\subsection{\label{sec:experiment} Statistical complexity is maximized between the synchronized and desynchronized cortical states}

By calculating the complexity of each  window, we verify that the complexity is low for both extreme situations (see Fig.~\ref{fig:raster}(e)): $C=0.190$ for very ordered (green circle) and $C=0.206$ for very disordered states (blue square) in Fig.~\ref{fig:raster}. 
Moreover, the complexity is higher for states with intermediate spiking variability: $C=0.304$ for the red triangle in Fig.~\ref{fig:raster}.

\begin{figure}[!ht]
\centerline{\includegraphics[width=0.75\linewidth]{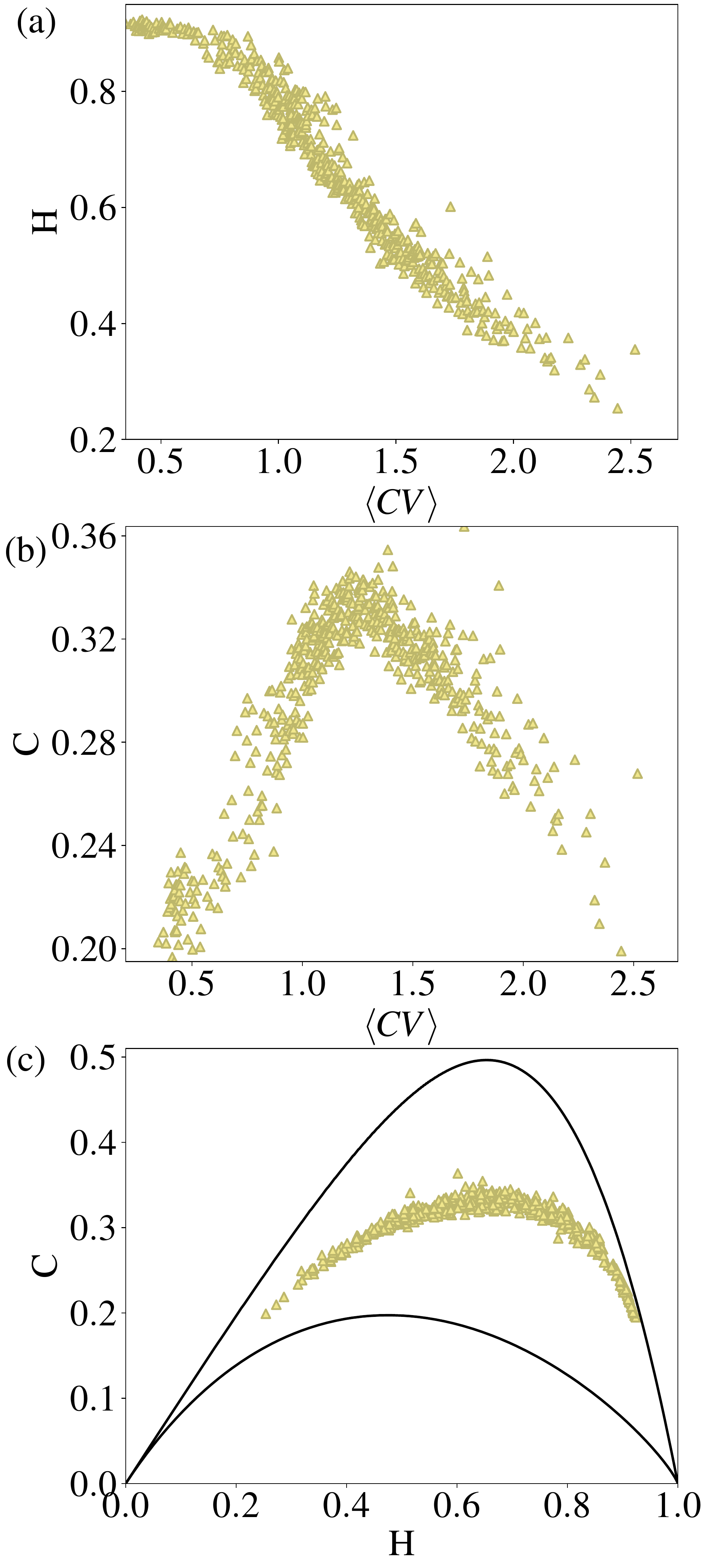}}
\caption{\label{fig:HCCV} 
Information-theory quantifiers for different levels of spiking variability of the cortical states of a single animal.
(a) symbolic Shannon entropy and
(b) statistical complexity versus coefficient of variation ($CV$) of the spiking activity. 
(c) Complexity$-$entropy plane. 
The peak in complexity occurs for $C_{max}=0.330\pm 0.001$, $\langle CV\rangle_{C_{max}}=1.31\pm 0.15$, $H_{C_{max}}=0.66 \pm 0.05$.}
\end{figure}

In Fig.~\ref{fig:HCCV}(b), we show an illustrative example of the 2D parameter space $C$ \textit{versus} $CV$ for all cortical states of a single rat. There is a clear peak of the complexity for intermediate values of the coefficient of variation of the population firing rate.
To find averages and standard errors for the statistical complexity across all windows and to determine confidence intervals for our metrics we grouped data by $CV$ in intervals of $0.15$ and by entropy $H$ in intervals of $0.05$. We also verified that these intervals are larger, therefore more conservative, than the ones obtained with the mean value of complexity and its standard error of the mean. 
For this rat we obtain a maximum value of complexity $C_{max}=0.330\pm 0.001$ for $CV_{C_{max}}=1.31\pm0.15$. 
The peak in the complexity can also be observed when plotted against the entropy, as shown in Fig.~\ref{fig:HCCV}(c). The peak occurs for $H_{C_{max}}=0.66\pm 0.05$.

\begin{figure}[t]
\subfigure{\includegraphics[width=0.75\columnwidth]{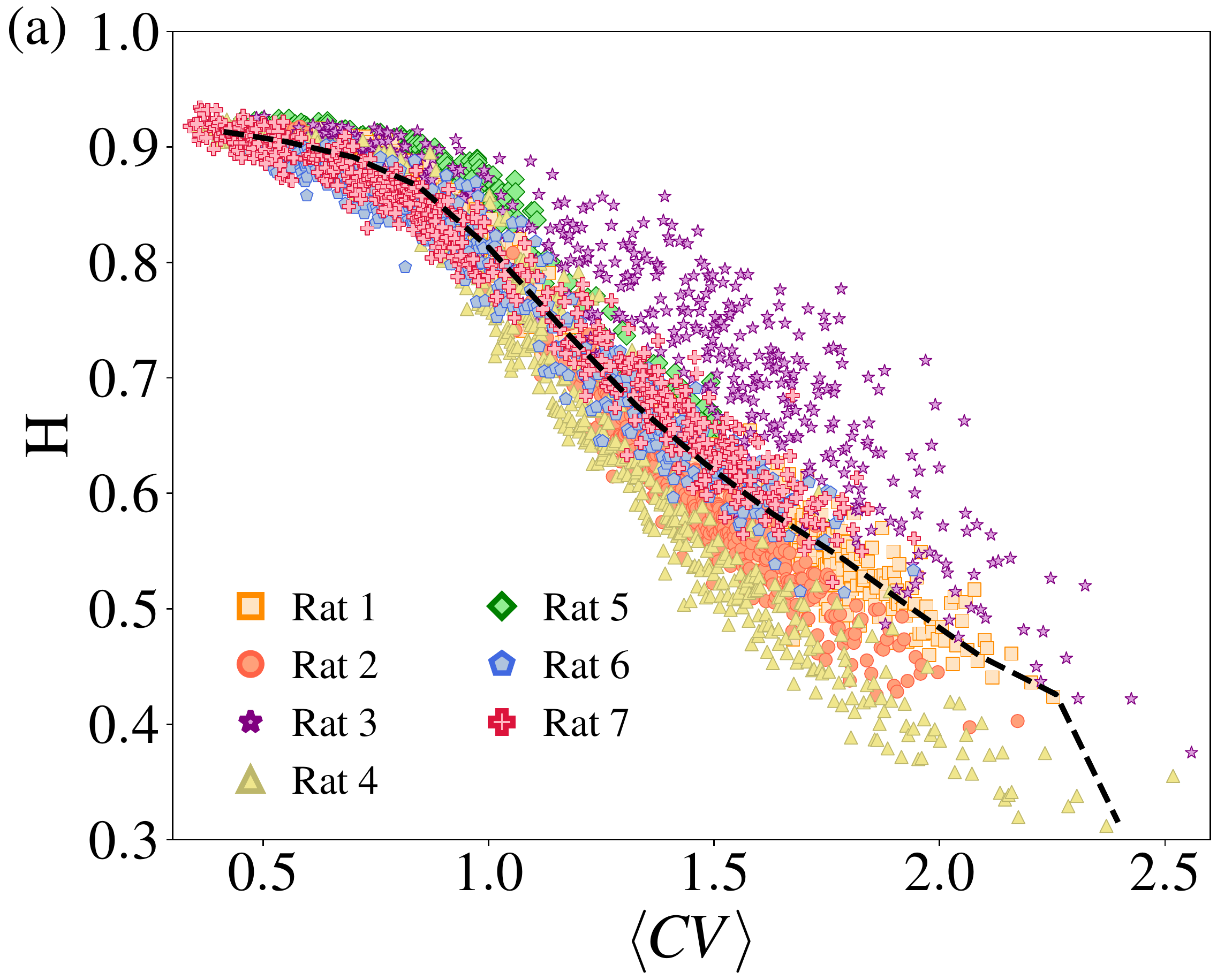}
}
\subfigure{\includegraphics[width=0.75\columnwidth]{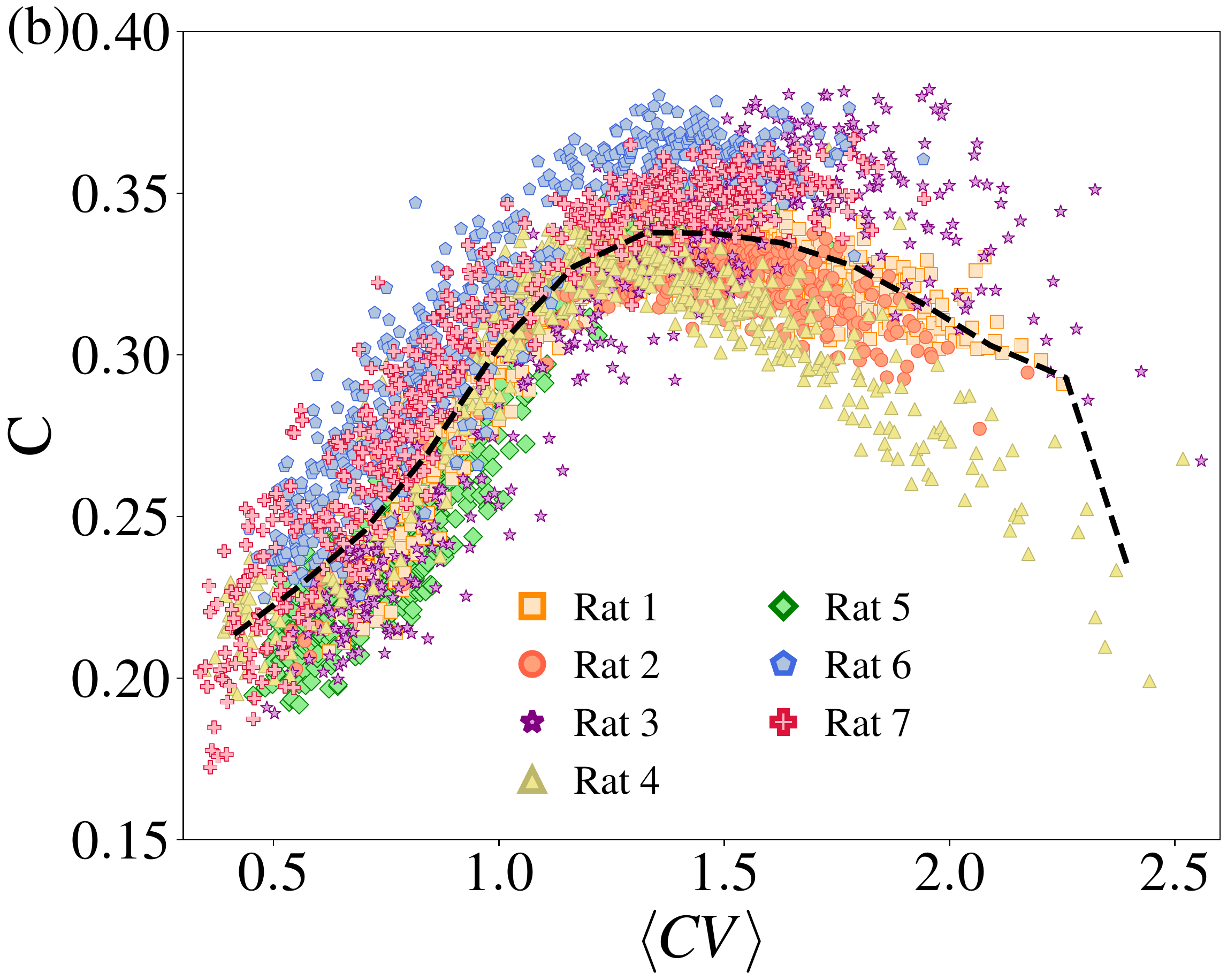}
}
\subfigure{\includegraphics[width=0.75\columnwidth]{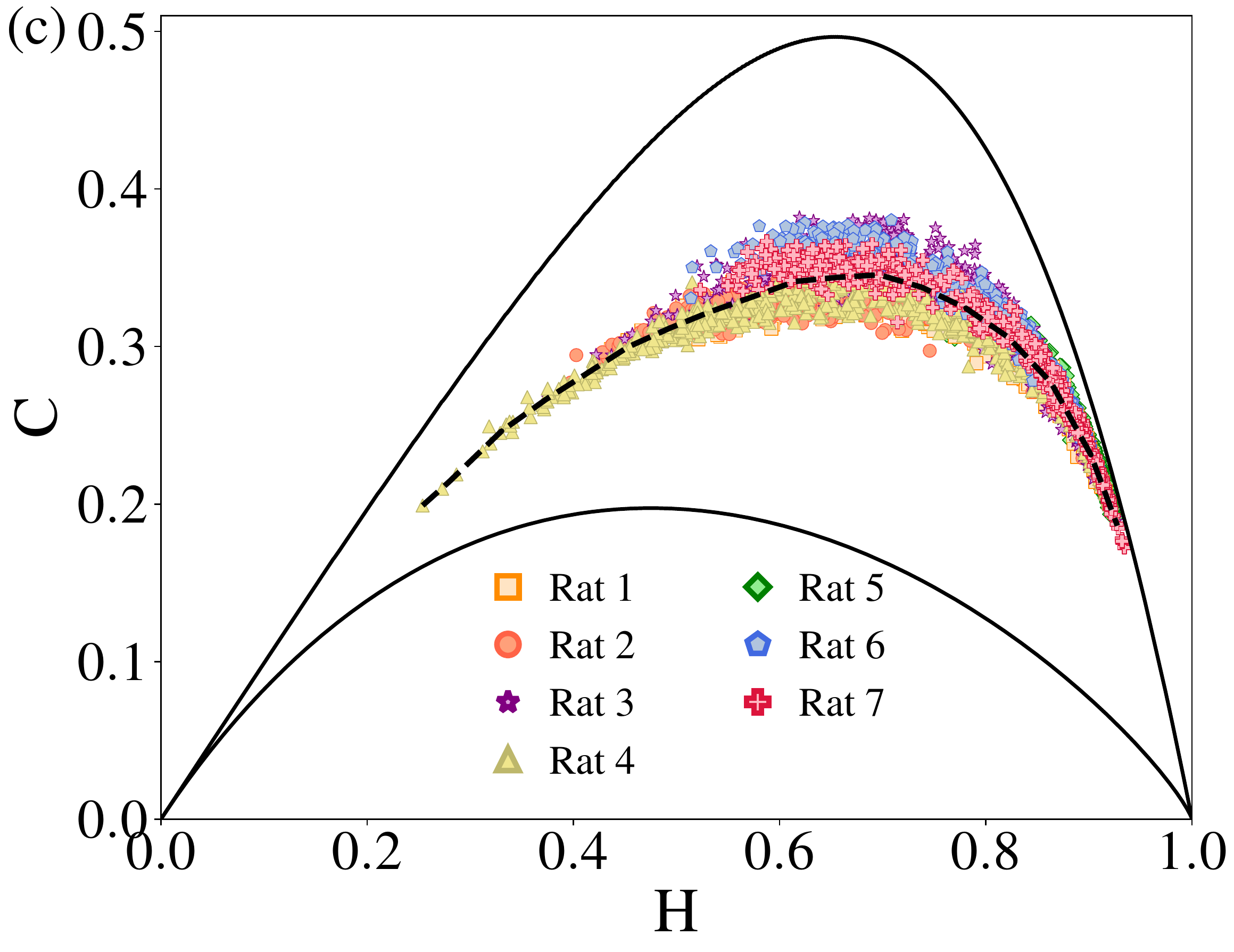}
}
\caption{\label{fig:allrats} 
Information-theory quantifiers as in Fig.~\ref{fig:HCCV} for all animals together.
(a) Shannon entropy and
(b) statistical complexity plotted along with the respective coefficient of variation ($CV$) of the spiking activity for each window of time. 
(c) Complexity$-$entropy plane. 
The dashed line represents the mean values of $C$ and $H$. 
The peak in complexity occurs for $C_{max}=0.333\pm 0.001$, $\langle CV\rangle_{C_{max}}=1.35\pm 0.15$, $H_{C_{max}}=0.70 \pm 0.05$}

\end{figure}

The group result obtained for all animals is shown in Fig.~\ref{fig:allrats}. Qualitatively, results are the same as for a single rat: inverse relation between $H$ and $CV$ (Fig.~\ref{fig:allrats}a) and a peak of complexity when compared to both $CV$ (Fig.~\ref{fig:allrats}b) and $H$ (Fig.~\ref{fig:allrats}c). 
At the peak, the mean values of our metrics for group data are:
$C_{max}=0.333\pm 0.001$, $CV_{C_{max}}=1.35\pm0.15$, $H_{C_{max}}=0.70\pm 0.05$.
The value of spiking variability $CV_{C_{max}}$ at which complexity is maximized agrees (within error bars) with the  values where  previous studies have found signatures of criticality~\cite{Fontenele19,lotfi2020signatures}.

The multi-scale complexity causality plane has been employed before to study brain signals~\cite{Montani15,Montani2014,Montani2015neuronas}. 
For simulated data, the causal $H \times C$-plane can be employed to separate more chaotic dynamics from more stochastic behavior~\cite{Rosso07,Zunino12}. 
Typically, one can explore the time scales to find the maximum values of complexity in a data set. Here we can explore different cortical states and determine the states with maximal complexity. 
We show that the complexity is maximized for cortical states with intermediate values of entropy (see Fig.~\ref{fig:HCCV}(c) and Fig.~\ref{fig:allrats}(c)).

\begin{figure}[t]
\subfigure{\includegraphics[width=0.8\columnwidth]{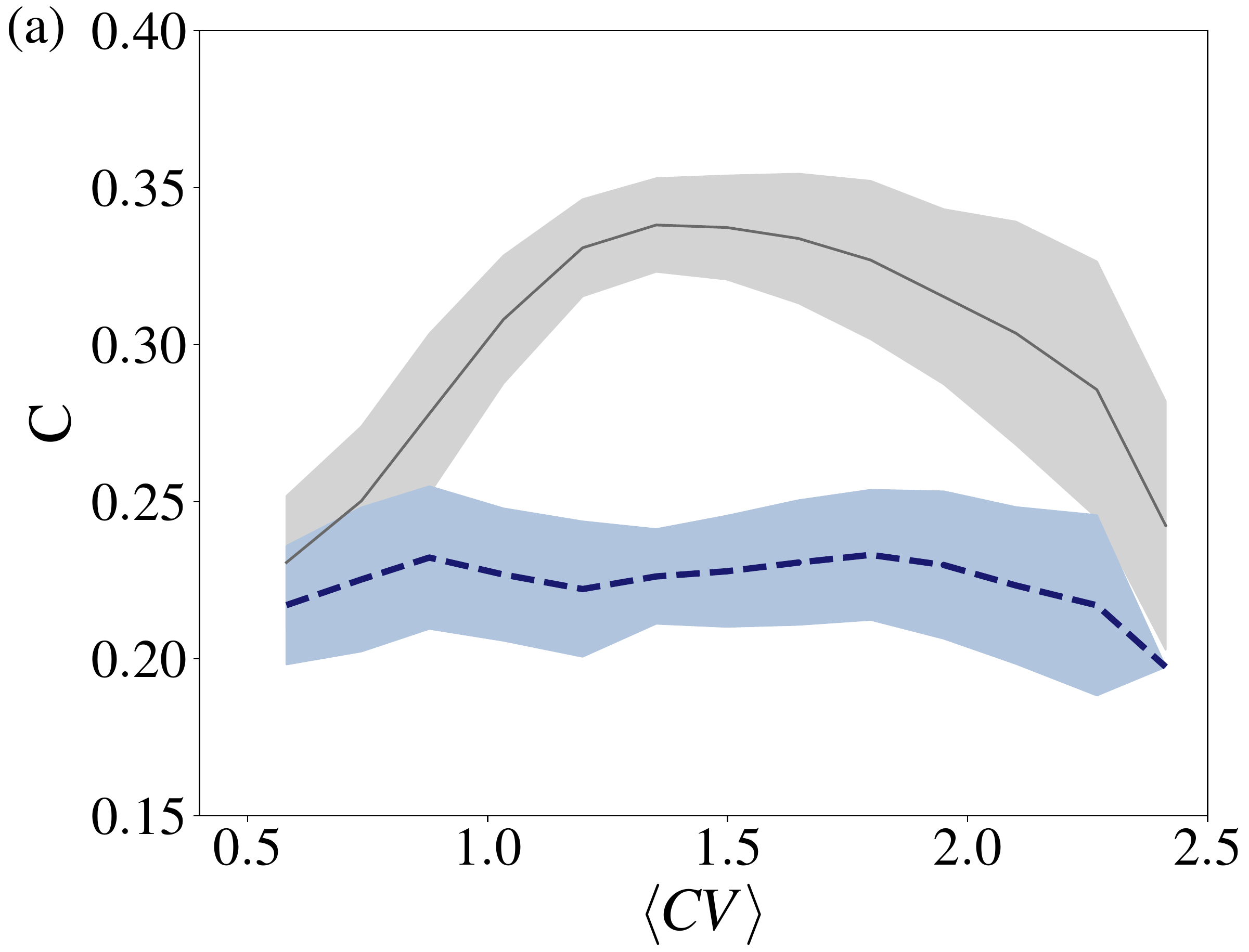}
}
\subfigure{\includegraphics[width=0.8\columnwidth]{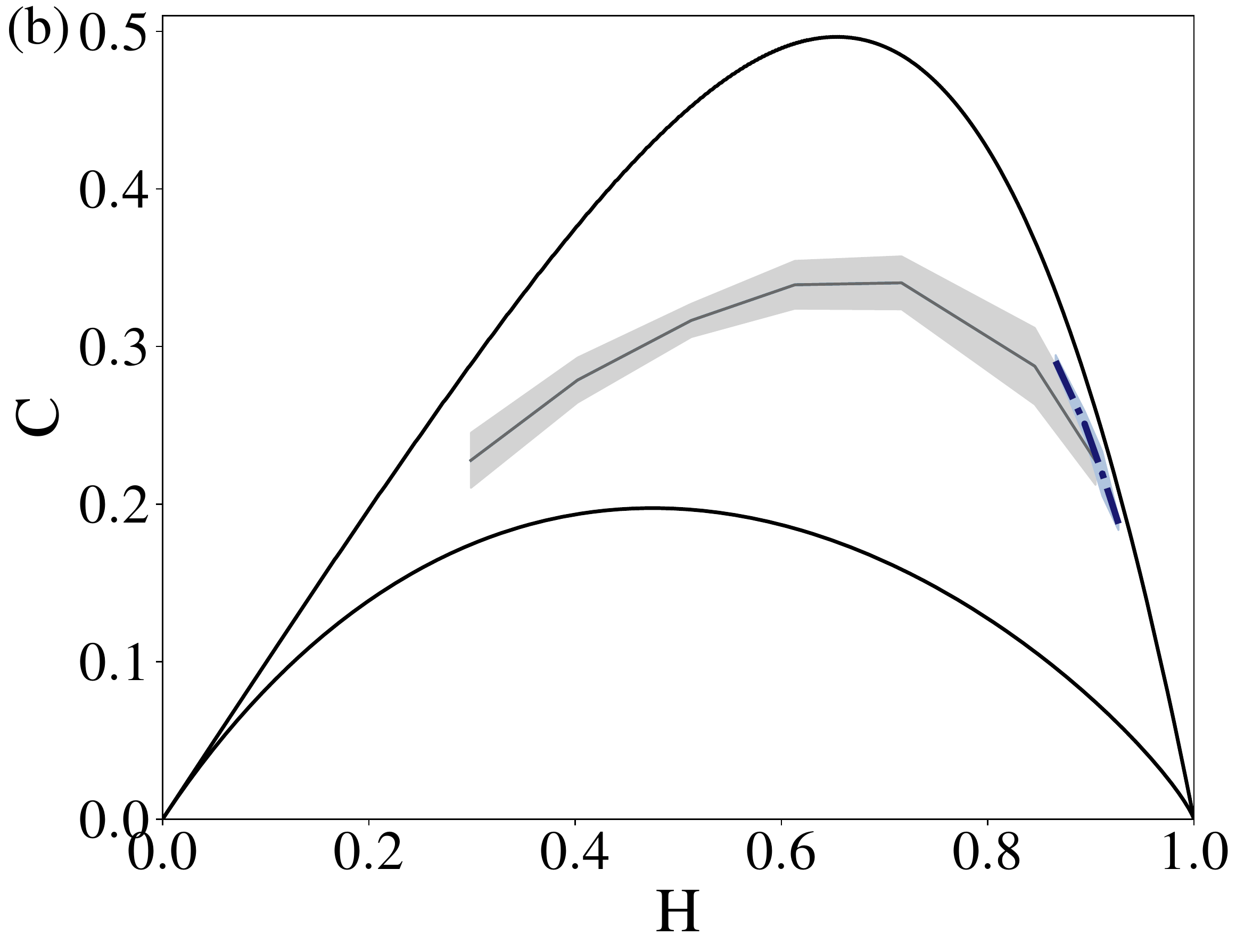}
}
\caption{\label{fig:allratsshuffle} Comparison between shuffled (dashed line) and real (continuous line) data for all rats. 
(a) Average (lines) and standard deviation (shading) of complexity $C$ plotted against the CV associated with each window before the shuffling. (b)  Average (lines) and standard deviation (shading) of complexity and entropy in the complexity-entropy plane.}
\end{figure}

The average values of the entropy and the complexity for all animals and their standard error can be compared with the same quantifiers for the randomized time series (Fig.~\ref{fig:allratsshuffle}). Results are not reproduced by shuffled data. 
We have calculated the complexity for the randomized window, and plot it against the original $CV$ values~(Fig.~\ref{fig:allratsshuffle}(a)). It is clear that the complexity is smaller for random sequences of firing rates.
The entropy, on the other hand, is maximum for the shuffled data (see Fig.~\ref{fig:allratsshuffle}(b)).

\subsection{Statistical complexity is maximized close to 
the phase-transition in a network model}

We extended our analysis of a probabilistic cellular automata network model~\cite{kinouchi2006optimal} in which the critical point is well defined. Each site $i$ in the network has five possible states: the resting state ($s_i = 0$); the excited state ($s_i = 1$), which represents the moment when the neuron fires an action potential; and three refractory states ($s_i = 2, 3, 4$) when the neuron cannot fire a spike. The $N$ sites are distributed in a random graph, where each site is randomly connected with $K$ other presynaptic sites. Connections are kept unchanged throughout the simulation (quenched disorder).

A site $i$ can go from the resting state to the excited state ($s_i(t) = 0 \to s_i(t+1) = 1$) in two ways: 1) it can be activated by an external stimulus, modeled here by a Poisson process ($p_h = 1 - \exp(-r\delta t)$); 2) the site $i$ can be activated, with probability $ p_ {ij} $, if a presynaptic neighbor $j$ is active at time $t$. The remaining transitions ($1 \to 2, \cdots , 4 \to 0$) happen with probability 1. The time step of the model corresponds to $\delta t = 1$~ms.
This model is branching process-like. We define the branching ratio $ \sigma = K \langle p_{ij} \rangle$, and by construction, we define $p_{ij}$ to be a random variable with uniform distribution in the interval [0, $2 \sigma / K$] so that $\sigma$ is the control parameter of our simulations.

In the absence of external stimulus ($r = 0$), the system undergoes a mean-field directed percolation (MF-DP) phase transition at $\sigma_c = 1$~\cite{kinouchi2006optimal}. For $\sigma <1$, any initiated activity will eventually die and the system always goes to the absorbent state ($s_i=0, \forall i$) which represents the subcritical regime. For $\sigma> 1$, any started activity will be self-sustaining and will continue indefinitely through the network, characterizing the supercritical regime.

In our simulations, we use $N = 10^5$ sites and each one has $K = 10$ presynaptic neighbors. We varied $\sigma$ around the critical point ($0.996 \leq \sigma \leq 1.010$) and using $r = 10^{-6}~\mathrm{ms^{-1}}$ which generates a very small external stimulus that supports starting new activities in the subcritical regime. We run our simulations for $10^7$ time steps which are compatible with three hours of recordings in the experiment. We analyze the firing rate time series of 100 randomly selected neurons in the network to calculate $CV$, $H$, and $C$ as described before.

\begin{figure}[t]
\subfigure{\includegraphics[width=0.8\columnwidth]{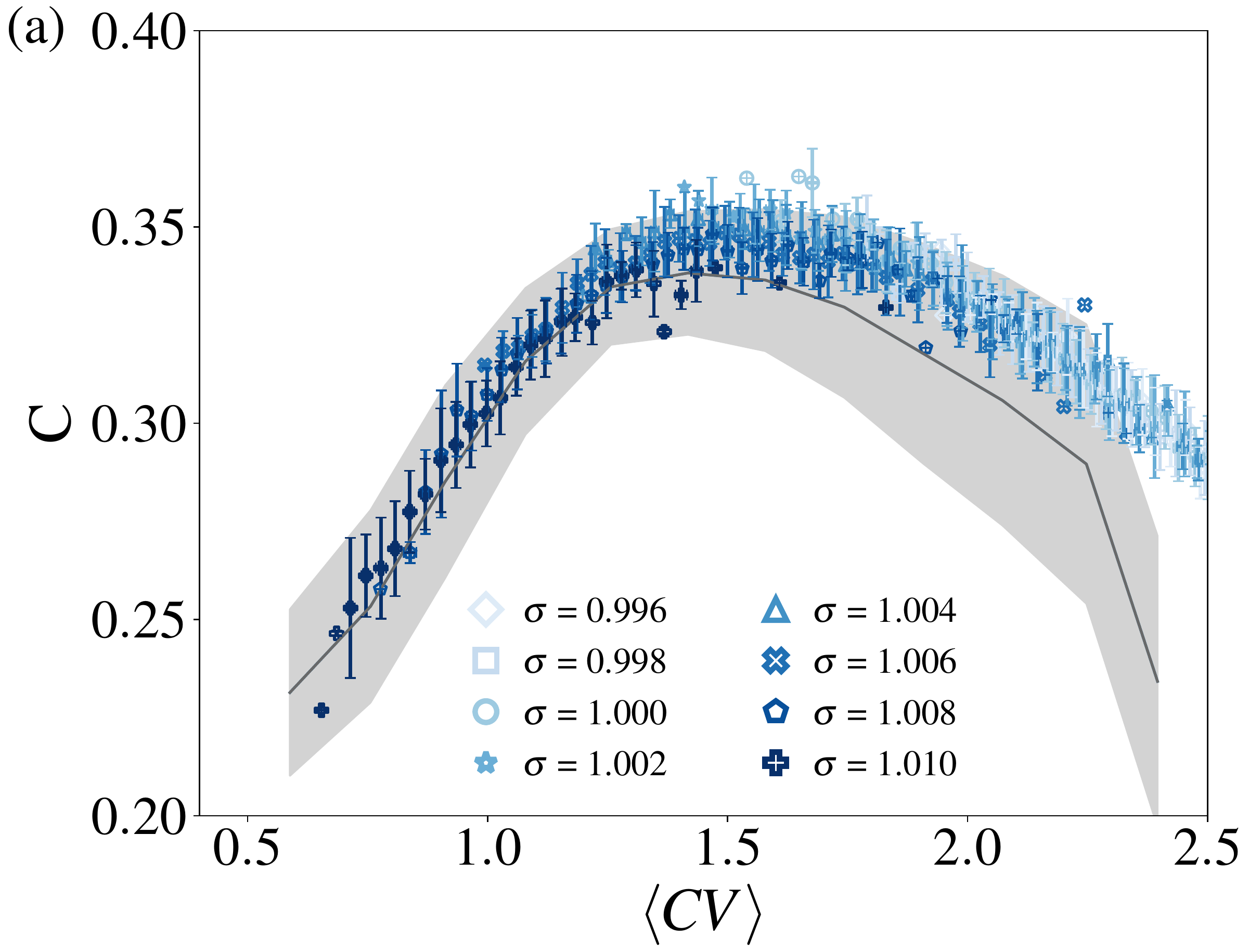}
}
\subfigure{\includegraphics[width=0.8\columnwidth]{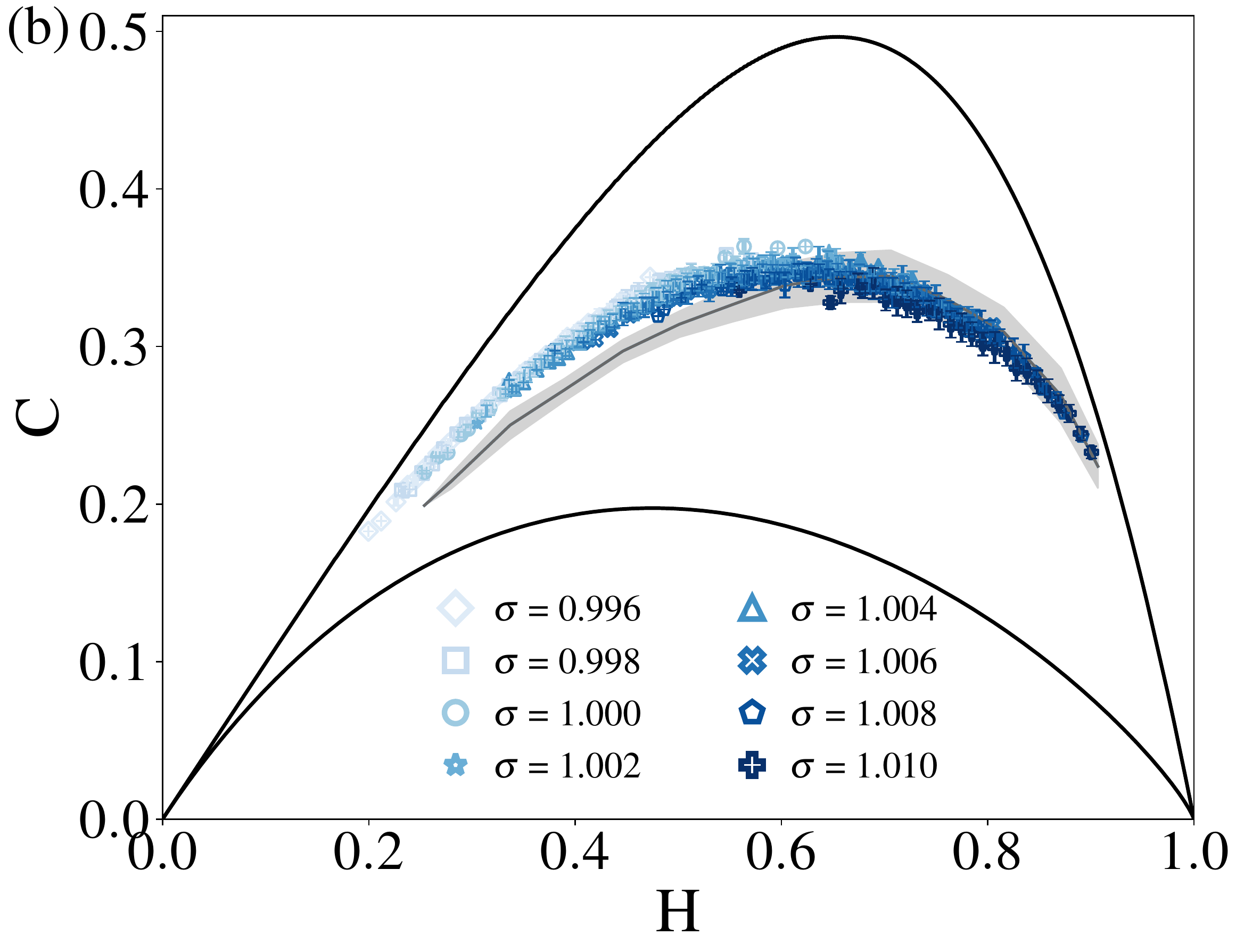}
}
\caption{\label{fig:model}
Complexity is maximized close to the criticality of a theoretical model. Comparison between model (blue dots) and group data with all animals (continuous line, same as in Fig.~\ref{fig:allratsshuffle}). Firing rates were analyzed with the same protocol in both cases. A mean-field directed percolation phase transition occurs for $\sigma = 1$ in the model~\cite{kinouchi2006optimal}.
(a) ($CV$,$C$) 2D projection space.
(b) Complexity$-$entropy plane. 
}

\end{figure}

In Fig.~\ref{fig:model}, we show the $C$ versus $CV$ diagram and the complexity-entropy plane to compare experimental group data and model. 
For the model, when analysed like the data, the complexity is maximized at $\langle CV\rangle_{C_{max}}= 1.58\pm0.15$, $H_{C_{max}}=0.60\pm0.05$ and its maximum value is $C_{max} = 0.347\pm0.001$.
We can cover the full extent of the experimental results by slightly varying the parameter $\sigma$ around its critical value. This result corroborates the claim that the urethane anesthetized cortex is operating near criticality\cite{Tawan2020,Beggs2014quasicriticality}.

\section{\label{conclusions}Concluding remarks}

To summarize, we have shown that cortical states can be characterized by information-theory quantifiers: Shannon permutation entropy~\cite{Bandt02} and Mart\'in-Platino-Rosso statistical complexity~\cite{Martin06}. 
Complexity is calculated from the Jensen disequilibrium measure and tends to zero for both regular and random signals. 
We have employed a symbolic representation of the firing rates (based on the Bandt-Pompe~\cite{Bandt02} recipe) to assign a probability distribution function to the time series generated by urethane-anesthetized rats and a simulated model~\cite{kinouchi2006optimal}. 

We have also shown that complexity in the population dynamics is maximized around $CV$ values where criticality signatures had been independently found via scaling analysis of neuronal avalanches~\cite{Fontenele19} and a maximum entropy approach~\cite{lotfi2020signatures}. 
Furthermore, we have shown that the experimental results were reproduced by a model.
In other words, the complexity is maximum close to the well defined critical point of the probabilistic cellular automata network model~\cite{kinouchi2006optimal}. These findings corroborate the results relating to the complexity and the criticality reported before in the neuronal model and culture~\cite{Timme16}.

Our study also opens new possibilities in investigating the complexity in the cortical states. First, we could characterize how the complexity changes during different cognitive tasks by analyzing neuronal firing rates (instead of local field potentials, for example~\cite{Montani15}). Second, we could relate the complexity and the criticality in awake animals, which would also allow us to compare the relationship among the complexity, the criticality, and the behavior.


\begin{acknowledgments}
The authors thank FAPEAL, UFAL, CNPq (grant 432429/2016-6) and CAPES (grant 88881.120309/2016-01) for financial support. NL is thankful to FACEPE (Grant No. BCT-0426-1.05/18) and CAPES (Grant No. 88887.308754/2018-00) for their support.
MC and PVC acknowledge support from CAPES (PROEX 534/2018 Grant No. 23038.003382/2018-39), FACEPE (Grant No. APQ-0642-1.05/18),  CNPq (Grants No. 301744/2018-1 and No. 425329/2018-6), and Universidade Federal de Pernambuco (UFPE).
\end{acknowledgments}
%

\bibliography{matias}

\end{document}